\documentclass[runningheads]{llncs}
\usepackage{graphicx}
\usepackage{makecell}
\usepackage{pbox}
\usepackage{float}

\begin{document}
\title{A Catalogue of Concerns for Specifying Machine Learning-Enabled Systems}
\author{Hugo Villamizar \and Marcos Kalinowski \and Hélio Lopes}
\authorrunning{Villamizar. Hugo et al.}
\institute{Pontifical Catholic University of Rio de Janeiro, Rio de Janeiro, Brazil\\
\email{\{hvillamizar, kalinowski, lopes\}@inf.puc-rio.br}}
\maketitle              % typeset the header of the contribution
\begin{abstract}

Requirements engineering (RE) activities for machine learning (ML) are not well-established and researched in the literature. Many issues and challenges exist when specifying, designing, and developing ML-enabled systems. Adding more focus on RE for ML can help to develop more reliable ML-enabled systems. Based on insights collected from previous work and industrial experiences, we propose a catalogue of 45 concerns to be considered when specifying ML-enabled systems, covering five different perspectives we identified as relevant for such systems: objectives, user experience, infrastructure, model, and data. Examples of such concerns include the execution engine and telemetry for the infrastructure perspective, and explainability and reproducibility for the model perspective. We conducted a focus group session with eight software professionals with experience developing ML-enabled systems to validate the importance, quality and feasibility of using our catalogue. The feedback allowed us to improve the catalogue and confirmed its practical relevance. The main research contribution of this work consists in providing a validated set of concerns grouped into perspectives that can be used by requirements engineers to support the specification of ML-enabled systems.

\end{abstract}

\section{Introduction}
\label{sec:introduction}

The incorporation of machine learning (ML) components into the software of companies from all sectors is increasingly common. We refer to these as ML-enabled systems. Ensuring the quality of such systems is essential to understand and evaluate their results. However, this is not an easy task. For instance, a good learning system is one in which the learning evolves and improves over time, the model creation is reproducible and maintainable, the users are aware of how often the predictions are right and wrong, and the customer knows how much ML helps the system to achieve their goals. 

Typically, when practitioners evaluate ML models they look only at measures such as accuracy, precision and recall. However, it is important to understand the big picture of the constraints these systems put on the overall development. Where will the model be executed? What data will it have access to? How fast does it need to be? What is the business impact of a false positive? A false negative? How should the model be tuned to maximize business results? Moreover, the model is just a component of a system as a whole. In fact, there are other components that require attention such as the infrastructure to deploy, update, and serve the model, the integration of the model with the rest of the system functionality and a user interaction design to build better experiences of using ML-enabled systems. 

Literature has shown that there are problems and challenges in the development of ML-enabled systems~\cite{arpteg2018software}\cite{lwakatare2019taxonomy}\cite{de2019understanding}\cite{wan2019does}. Requirements engineering (RE) is no stranger to this. The impact of incomplete/hidden requirements on overall system development is considered one of the most critical problems of RE in practice ~\cite{mendez2017naming}, and the difficulty to specify complete requirements increases for ML-enabled systems, where requirements engineers are typically not aware of the concerns that should be considered for such specification. Some recent research papers have drawn the attention of researchers and practitioners on the fact that ML can benefit from RE ~\cite{cysneiros2020non}\cite{dalpiaz2020requirements}\cite{habibullah2021non}\cite{heyn2021requirement}. However, current research on the intersection between RE and ML mainly focuses on using ML techniques to support RE activities rather than on extending existing requirements techniques or providing new ones to support ML-enabled systems. In line with this, the roadmap for the future of SE proposed by the Carnegie Mellon University Software Engineering Institute~\cite{carleton2021architecting} emphasizes that existing RE methods will need to be expanded to decouple ML problem and model specification from the system specification. Indeed, recent literature reviews~\cite{ahmad2021s}\cite{villamizar2021requirements} show that topics such as identifying quality attributes, specifying them, and understanding how they can be analyzed are not well-established and researched in the context of ML. 

%% This landscape showed us there is an incredible amount of work to be done between the definition of requirements for ML, the customer expectations, the development of a model and the incorporation of it into a system.

RE can improve the development of ML-enabled systems in several ways. For example, by identifying quality metrics beyond accuracy, dealing better with human factors and understanding why models do not fit and for whom they do not fit. With the aim at addressing one of the main problems presented in current RE for ML research, in this work we propose a catalogue of 45 concerns that can be used by requirements engineers to support the specification of ML-enabled systems. Based on our recent systematic mapping study ~\cite{villamizar2021requirements}, on insights from industrial experiences~\cite{hulten2019building}\cite{kalinowski2020lean} we propose a catalogue organized into five different perspectives: objectives, user experience, infrastructure, model, and data. To validate our catalogue of concerns we conducted a focus group session with eight software professionals experienced with developing ML-enabled systems. The results revealed that the professionals were not explicitly aware of many of the concerns, but that they recognized their relevance and potential impact on the overall system being developed. In general, they agreed with the concerns and the way of grouping them into perspectives. In addition, we received relevant feedback that we used to improve our catalogue.

The remainder of this paper is organized as follows. Section~\ref{sec:background} provides the background and presents related work. Section~\ref{sec:methodology} describes the research methodology. Section~\ref{sec:catalogue} presents our catalogue of concerns for specifying ML-enabled systems. Section~\ref{sec:focus_group} presents the focus group session we conducted and how it contributed to our catalogue. We discuss and conclude our work in Section~\ref{sec:discussion} and Section~\ref{sec:conclusions}, respectively.

\section{Background and Related Work}
\label{sec:background}

This section provides a background on RE for ML-enabled systems, presenting particularities that make RE essential in this context. Related work is also presented.

\subsection{RE for ML-enabled Systems} 
\label{subsec:re_for_ml}

ML is the study of computer algorithms that explores data to determine the best way to combine the information contained in the representation (training data) into a model that generalizes to data it has not already seen~\cite{mitchell1997machine}. This type of systems, unlike traditional software systems, base its behavior on data from the external world instead of explicitly programming hard rules. In other words, data, to some extent, replace code. However, data may not be adequate and lead to bad outcomes. The output of a model is a prediction, sometimes surprisingly accurate and sometimes surprisingly inaccurate. This supposes a change in the way of designing, developing and testing ML-enabled systems.

RE and ML have a special connection. We can see a machine-learned model as a specification based on training data since data is a learned description of how the model shall behave~\cite{kastner2020machine}. Nevertheless, most ML models lack requirements specifications since current RE practices are not well defined and organized for ML~\cite{ishikawa2019engineers}\cite{kuwajima2020engineering}. Assuring the quality of the specifications is crucial for project success since misunderstandings and defects in requirements documents can easily lead to design flaws and cause severe and costly problems. Good requirements should be, at a minimum, complete, consistent, correct, unambiguous and testable~\cite{shull1998developing}.

There are a considerable number of concerns when designing ML-enabled systems. Given their inherent nature, requirements for these systems must have a stronger focus on data requirements~\cite{challa2020faulty}. For instance, Vogelsang~\cite{vogelsang2019requirements} based on interviews with data scientists, mentioned that requirements engineer has to identify and specify requirements regarding the quantity, quality, collection, formats and the ranges of data. This with the aim of analyzing the given dataset against requirements and business goals. Furthermore, it is well known that in ML projects, model performance measures must be specified. These measures, to some extent, depend on the quality and pre-processing of the data. However, the importance of specifying these measures in a way that customers can understand and analyze it to make decisions is often overlooked~\cite{de2019understanding}. Another concern in requirements for ML is the definition of quality properties, also known as non-functional requirements (NFRs)~\cite{cysneiros2020non}~\cite{horkoff2019non}. For instance, depending on the context of classification problems, the requirements engineer might have to specify ethical concerns, particularly if it is people who are classified, and define what characteristics should not be used for classification~\cite{aydemir2018roadmap}. Similarly, other quality properties such as explainability, transparency, modularity, testability, security and privacy should be clearly specified~\cite{cysneiros2018software}. 

RE practices are not well established for ML-enabled systems. Some secondary studies have recently given attention to this~\cite{ahmad2021s}\cite{villamizar2021requirements}. For instance, while there is still limited research, the SE community agrees on the need of extending RE techniques for ML. Furthermore, the new proposed techniques have not yet been applied in practice~\cite{shin2019data}. On the other hand, there is a misconception about the new role of data scientists that often are taking of responsibilities such as defining and eliciting requirements by themselves~\cite{kim2017data}. Another particular challenge is the overconfidence of using ML~\cite{dimatteo2020requirements}. Customers see ML as magic, in other words, “ML will solve everything”. It's the job of RE to let stakeholders know the limitations and manage their expectations.

\subsection{Related Work}
\label{subsec:Related_work}

The intersection of RE and ML has been studied in recent years by the RE community and discussed in renown SE conferences~\cite{dalpiaz2020requirements}\cite{habibullah2021non}\cite{heyn2021requirement}. Several studies have investigated what properties or components should be considered by requirements engineers when designing ML-enabled systems. Here, we focus on an overview of work we consider directly related to our research.

Chuprina et al.~\cite{chuprina2021towards} present an artefact-based RE approach for the development of data-centric systems. The proposal encompasses four layers: context, requirements, system, and data, where each contains a set of concerns. We consider this work strongly related to our research. However, we found some differences. Firstly, their scope concerns data-centric systems, while ours is specifically related to machine learning. Indeed, our intention is to be more specific, including more fine-grained concerns that can be easily considered by practitioners in the ML-enabled system context. In addition, we detail ML-related concerns that we faced in practice that were not considered as part of their proposal, such as concerns related to user experience and infrastructure, which in our context showed being important for the success of ML-enabled systems.

Nakamichi et al.~\cite{nakamichi2020requirements} propose a requirements-driven model to determine the quality properties of ML systems. They list a set of issues related to ML. The authors cover perspectives such as environment/user, system/infrastructure, model, data and quality characteristics. In line with~\cite{chuprina2021towards}, the study is a great contribution in the field, however, we believe that our catalogue goes one step further than their described issues, by mapping additional perspectives and relevant concerns, for instance related to user experiences and ML objectives.

Another study we consider relevant is one conducted by Nalchigar~\cite{nalchigar2021modeling}. They report on an empirical study that evaluates a conceptual modeling framework for ML solution development for the healthcare sector. It consists of three views consumed by business people, data scientists, and data engineers. We know that this contribution can help to address ML model concerns, however, we believe that other views such as infrastructure and user experience should be broken down into concerns to cover the full system perspective.

More recently, Berry~\cite{berry2022requirements} discussed some RE related ideas on how to use measures to evaluate AI related solutions from the point of view of recall and precision. Hence, he addresses what we identify as the model performance concern. However, practical experiences show many other relevant concerns that should be considered when designing ML-enabled systems (e.g., ethics, explainability, accountability).

\section{Methodology for Defining the Concerns}
\label{sec:methodology}

\begin{figure}[ht]
    \centering
    \includegraphics[width=1\textwidth]{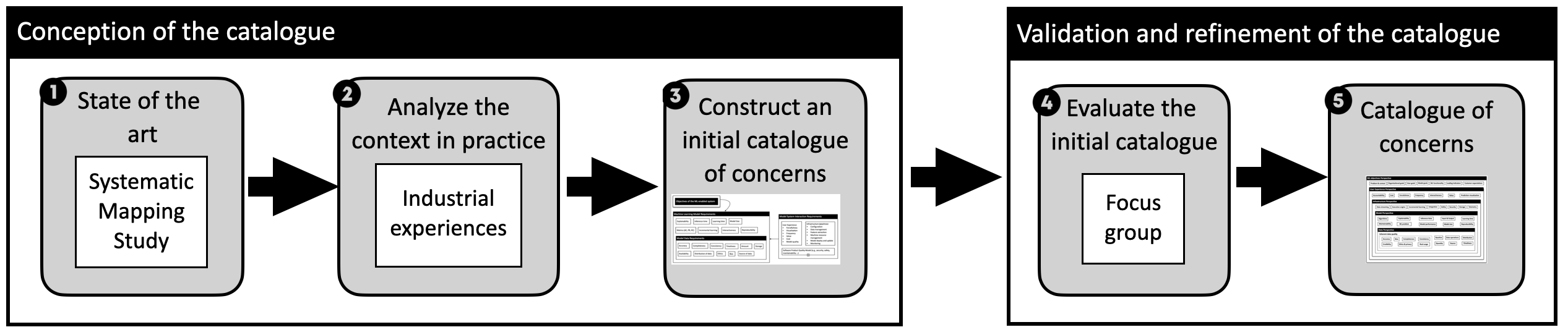}
    \caption{Overview of the study steps for defining the catalogue}
    \label{fig:conceiving_solution}
\end{figure}

We used the constructionism theory~\cite{fosnot2013constructivism} that advocates a person needs to understand how something works before exploring the different ways to construct solutions. Figure~\ref{fig:conceiving_solution} illustrates what we did and how we created, validated and improved our catalogue to support the specification of ML-enabled systems.

In order to understand what needs to be created, first, we conducted a literature review~\cite{villamizar2021requirements} on how ML could benefit from the RE perspective and what research opportunities could be addressed (step 1). The next step involved getting insights from industrial experiences (step 2). During the last two years, the first author has participated in R\&D projects involving the development of ML-enabled systems as part of the ExACTa~\footnote{http://www.exacta.inf.puc-rio.br} initiative~\cite{kalinowski2020lean}, which is co-coordinated by the second and the third authors. These projects involved several deliveries of solutions involving different types of machine learning problems and algorithms (e.g., decision trees, logistic regression, neural networks) to industrial partners. In experiential learning, this is defined as learning through reflection on doing. We took advantage of these planned synergies between theory and practice to learn more about the context in which the concerns operate. Additionally, we took advice from an industry-oriented publication based on more than a decade of experience building intelligent systems~\cite{hulten2019building}, with the aim at positioning the knowledge and insights acquired up to this point.

After analyzing the state of the art and the ML-enabled system development context in practice, we created the initial catalogue of concerns (step 3). It is noteworthy that the catalogue was critically reviewed by the second and the third authors, which are scientists active in the areas of software engineering and data science and that have been exploring the intersection between these two areas. The literature review led us to focus on requirements definition, since this was one of the main identified research challenges, and revealed several quality properties of ML-enabled systems. Our industrial experiences allowed us to validate our findings and revealed complementary perspectives and concerns to be considered for ML-enabled systems. Finally, we conducted a focus group session (step 4) with eight software professionals with large experience developing ML-enabled systems. The results of the focus group allowed us to adjust the initial proposal (step 5).

\section{Catalogue of Concerns for ML-Enabled Systems}
\label{sec:catalogue}

\begin{figure}
    \centering
    \includegraphics[width=0.83\textwidth]{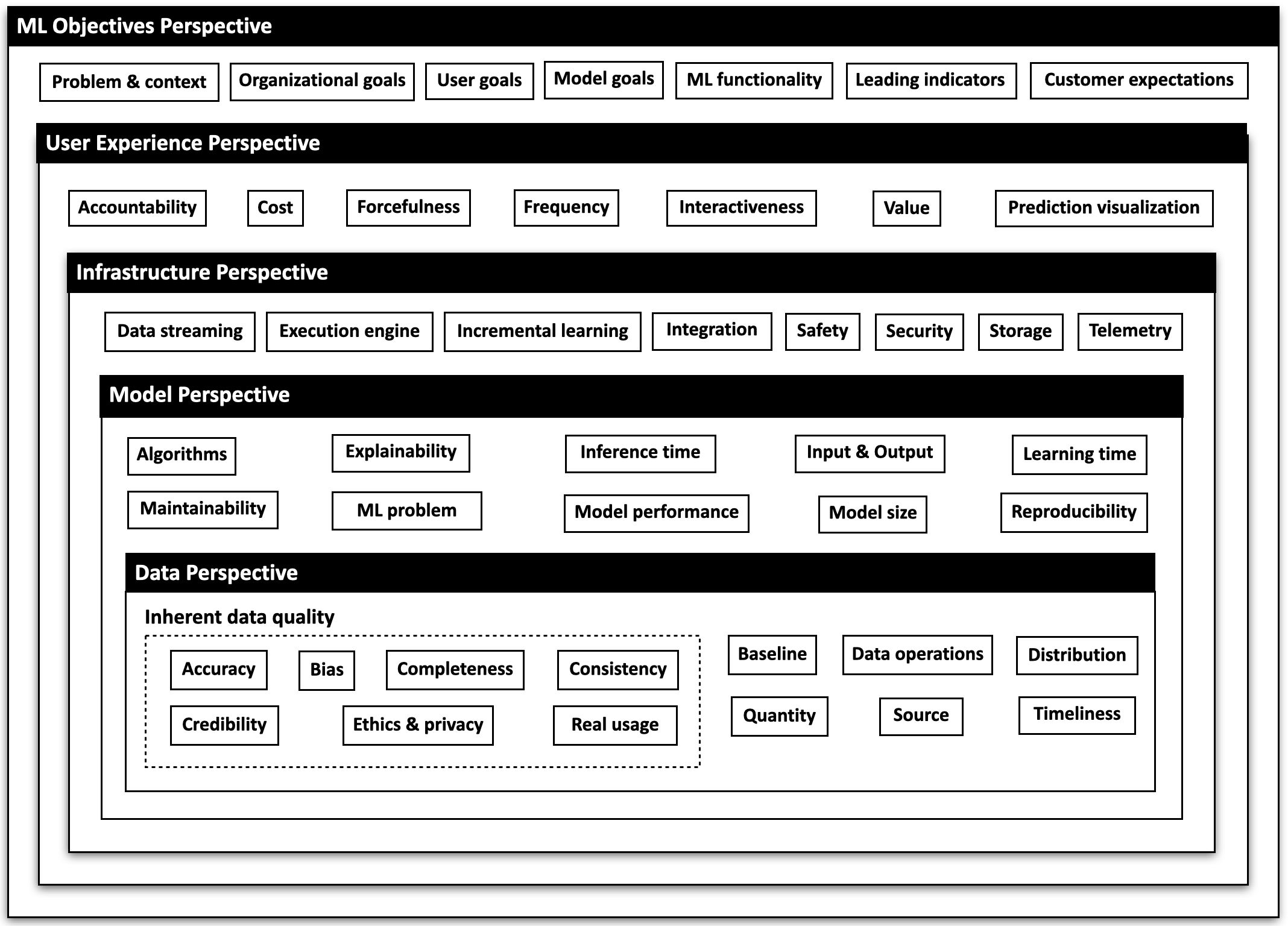}
    \caption{An overview of our catalogue: Perspectives and Concerns}
    \label{fig:overview_concerns}
\end{figure}

The success of ML-enabled systems is related to taking care of not only models and data, but also business context, user experience and infrastructure. The specification of ML-enabled systems involves concerns that are often not easily identified, resulting in hidden requirements. For instance, it is clear that a model needs good data to be trained and then evaluated, but it is not clear what are the criteria that define the data as good, nor that defining the frequency and forcefulness of the model is important to get better user experiences, nor that the model needs to be integrated with other services to ingest new data. Therefore, herein we propose a catalogue (Figure~\ref{fig:overview_concerns}) considering five complementary perspectives that accommodate the findings of our literature review and that showed being relevant in practice, building a big picture for specifying ML-enabled systems that has not previously been formalized.

The perspectives we identified and validated as relevant and complimentary are: objectives, user experience, infrastructure, model, and data. The depicted version already considers the adjustments made based on the focus group feedback. We believe that this catalog can be used by requirements engineers to support the specification of ML-enabled systems, making them aware of the big picture and helping to avoid incomplete/hidden requirements. We suggest the concerns to be analyzed by requirements engineers and discussed with stakeholders to understand the degree to which related requirements should be met. We describe the perspectives and detail their concerns hereafter, focusing on what the concerns mean in terms of RE specifications.

\subsubsection{ML objectives perspective:} 
\label{subsubsec:goals_measures_artifact}

Bridging the gap between the high-level goals and the detailed properties of ML models is one of the most common causes of failure to succeed~\cite{barash2019bridging}. Table~\ref{table:goal_properties} presents the concerns related to ML objectives that could influence the specification of ML-enabled systems. 

\begin{table}[H]
\scriptsize
\centering
\caption{ML objective perspective}
\label{table:goal_properties}
\begin{tabular}{|p{2.9cm}|p{9cm}|}
\hline
\multicolumn{1}{|c|}{\textbf{Concern}} & \multicolumn{1}{c|}{\textbf{Description (Addressing this concerns involves ... )}}\\ \hline

Problem \& Context                 &  
Specifying the problem that ML will address and its context before coding. ML must be targeted at the right problem.\\ \hline

%Understanding the business problem before coding is essential. ML must be targeted at the right problem.

Organizational goals     & 
Specifying measurable benefits ML is expected to bring to the organization. E.g., increase the revenue in X\%, increase the number of units sold in Y\%, number of trees saved. \\ \hline

%E.g., revenue, profit, number of units sold, trees saved, lives improved.

User goals               & 
Specifying what the users want to achieve by using ML. E.g., for recommendation systems this could involve helping users finding content they will enjoy.\\ \hline

%it is about setting goals around questions and decisions. E.g., if your system is about helping users find content they will enjoy, are they finding content that they end up liking?

Model goals       & 
Specifying metrics and acceptable measures the model should achieve (e.g., for classification problems this could involve accuracy $\geq$ X\%, precision $\geq$ Y\%, recall $\geq$ Z\%).  \\ \hline

%Metrics such as accuracy, precision, recall, F1-score.

ML functionality        & 
Specifying the ML results in terms of functionality that the model will provide (e.g., classify customers, predict probabilities). \\ \hline

%it defines the result of ML in terms of functionality (e.g., classify customers, predict probabilities).

Leading indicators       & 
Specifying measures correlating with future success, from the business' perspective. This could include the users' affective states when using the ML-enabled system (e.g., customer sentiment and engagement).  \\ \hline

%measure that correlates with future success. E.g., customer sentiment that measures how they feel about the system, and customer engagement that measures how much they use the system.

Customer expectations      & 
Specifying expectations of customers and end-user in terms of how the system should behave (e.g., how often they expect predictions to be right or wrong). \\ \hline

% users should understand how often the prediction is right and wrong.

\end{tabular}
\end{table}

\fontsize{9pt}{12pt}
\selectfont

\subsubsection{User experience perspective:} 
\label{subsubsec:user_exp_perspective}

Better ML includes building better experiences of using ML. Connecting the predictions with users is critical for achieving success. However, this is not trivial and requires deep analysis. For instance, one mistake in the way of interacting with the user, and all the work spent on data pre-processing and modeling may be wasted. The goal of this perspective is to create effective user experiences (UX) so that users can interact appropriately and understand what is going on with the predictions of the model. Table~\ref{table:user_exp_properties} shows the user experience concerns and a brief description of each one.

\begin{table}[H]
\scriptsize
\centering
\caption{User Experience Perspective.}
\label{table:user_exp_properties}
\begin{tabular}{|p{2.8cm}|p{9cm}|}
\hline
\multicolumn{1}{|c|}{\textbf{Concern}} & \multicolumn{1}{c|}{\textbf{Description (Addressing this concerns involves ... )}}\\ \hline

Accountability  & Specifying who is responsible for unexpected model results or actions taken based on unexpected model results. \\ \hline

%the users should be aware of who is responsible for unexpected model results.

Cost            & Specifying costs involved in executing the inferences and also the user impact of a wrong model prediction. \\ \hline

% users should be aware of the impact of a wrong prediction.

Forcefulness    & Specifying how strongly the system forces the user to do what the model indicates they should (e.g., automatic or assisted actions).\\ \hline

% how strongly the experience encourages the user to do what the model thinks they should. E.g., automatic or assisted actions.

Frequency       &  Specifying how often the system interacts with users (e.g., interact whenever the user asks for it or whenever the system thinks the user will respond). \\ \hline

% how often the experience tries to interact with users. E.g., interact whenever the user asks for it or whenever the system thinks the user will respond.

Interactiveness    & Specifying what interactions the users will have with the ML-enabled system, (e.g., to provide new data for learning, or human-in-the-loop systems where models require human interaction). \\ \hline

%the experience should allow interactions with users in order to get data to grow the learning.

Value           & Specifying the added value as perceived by users from the predictions to their work. \\ \hline

%how much the user thinks it benefits and how much it helps the system achieve its goals.

Prediction visualization    & Specifying how the ML outcomes will be presented so that users can understand them (e.g., specifying dashboard and visualization prototypes for validation). \\ \hline

%how the ML outcomes will appear so that users can understand it. E.g., organizing content, adding information to a display.

\end{tabular}
\end{table}

\subsubsection{Infrastructure perspective:} 
\label{subsubsec:infrastructure_perspective}

%This perspective covers the important components beyond the model, that need to be built, or at least considered, when we think about ML-enabled systems. Typically, data scientists are in charge of the entire life-cycle of ML projects~\cite{kim2017data}, which can result in software that does not focus on users needs and in a product with high technical debt~\cite{sculley2015hidden}. 

ML models need to be integrated with other services. This includes components such as ingesting and learning from new data. The adoption of ML is growing, but proper implementations are needed to fulfill their promise. Hence, software engineers play an important role to orchestrate these ML components. Table~\ref{table:infra_properties} presents the infrastructure concerns.

\begin{table}[H]
\scriptsize
\centering
\caption{Infrastructure Perspective.}
\label{table:infra_properties}
\begin{tabular}{|p{2.8cm}|p{9cm}|}
\hline
\multicolumn{1}{|c|}{\textbf{Concern}} & \multicolumn{1}{c|}{\textbf{Description (Addressing this concerns involves ... )}}\\ \hline

Data streaming          & Specifying what data steaming strategy will be used (e.g., real time data transportation or in batches). \\ \hline

% code that transports data from the source to a target for further processing. It can be in real time or batches.

Execution engine        & Specifying how the model of the ML-enabled system will be executed and consumed (e.g., client-side, back-end, cloud-based, web service end-point). \\ \hline

% code that defines how to consume the model (e.g., client-side, server-centric, web service, back-end).

Incremental learning    & Specifying the need for ML-enabled system abilities to continuously learn from new data, extending the existing model’s knowledge. \\ \hline

% code that allows input data to be continuously used to extend the existing model’s knowledge.

Integration             & Specifying the integration that the model will have with the rest of the system functionality.  \\ \hline

% there must be code integrating the model with the rest of the system functionality.

Safety       & Specifying how the system deals with risks to prevent dangerous failures. Critical systems that incorporate ML should analyze the probability of the occurrence of harm and its severity.\\\hline

% critical systems that incorporate ML should analyze the probability of the occurrence of harm and its severity.

Security     & Specifying how the system deals with security issues (e.g., vulnerabilities) to protect the data. ML systems often contain sensitive data that should be protected. \\ \hline

% ML systems often contain sensitive data that should be protected.

Storage                 & Specifying where the ML artifacts (e.g., models, data, scripts) will be stored. \\ \hline

%where the ML artifacts will be stored.

Telemetry               & Specifying what ML-enabled system data needs to be collected. Telemetry involves collecting data such as clicks on particular buttons and could involve other usage and performance monitoring data.  \\ \hline

%code for collecting data such as clicks on a particular button.

\end{tabular}
\end{table}

\subsubsection{Model perspective:} 
\label{subsubsec:model_perspective}

Building a model implies not only training an algorithm with data to predict or classify well some phenomenon. Many other aspects determine its success. The aim of this perspective is to provide a set of model concerns a requirements engineer should analyze. Table~\ref{table:model_properties} presents these concerns. 

\begin{table}[H]
\scriptsize
\centering
\caption{Model Perspective.}
\label{table:model_properties}
\begin{tabular}{|p{2.8cm}|p{9cm}|}
\hline
\multicolumn{1}{|c|}{\textbf{Concern}} & \multicolumn{1}{c|}{\textbf{Description (Addressing this concerns involves ... )}}\\ \hline

Algorithms     & Specifying the set of algorithms that could be used/investigated, based on the ML problem and other concerns to be considered (e.g., constraints regarding explainability or model performance, for instance, can limit the solution options). \\ \hline

% Based on the problem, data scientist uses several algorithms.

Explainability       & Specifying the need to understand reasons of the model inferences. The model might need to be able to summarize the reasons of its decisions. Other related concerns, such as transparency and interpretability, may apply. \\ \hline

%models that are able to summarize the reasons of their decisions. Other concerns such as transparency and interpretability apply.

Inference time       & Specifying the acceptable time to execute the model and return the predictions.\\ \hline

% time needed to execute the model and returns the predictions.

Input \& Output       & Specifying the expected inputs (features) and outcomes of the model. Of course, the set of meaningful inputs can be refined/improved during pre-processing activities, such as feature selection. \\ \hline

% the data of the model must contain the context of the problem to be solved (e.g., features, outcomes, labels).

Learning time        & Specifying the acceptable time to train the model.\\ \hline

% time needed to train the model.

Maintainability   & Specifying the need for preparing the model to go through changes with reasonable effort (e.g., refactoring, documentation, automated redeployment). \\ \hline

% E.g., best coding practices, refactoring and documentation apply.

ML problem type & Specifying the problem type tackled by the ML algorithm (e.g., classification, regression, clustering, extract information from text).\\ \hline

% predict values or categories, discover structure, find unusual occurrences and extract information from text.

Model performance  & 
Specifying the metrics used to evaluate the model (e.g., precision, recall, F1-score, mean square error) and measurable performance expectations. \\ \hline

% there are different types of metrics to evaluate models (e.g., accuracy, precision, recall, F1 score, mean square error, log loss).

Model size           & Specifying the size of the model in terms of storage and its complexity (e.g., for decision trees there might be needs for pruning).\\ \hline

%size of the model in terms of storage and model complexity.

Reproducibility      & Specifying the need for replicating the model creation process and its experiments. \\ \hline

% Building pipelines help to improve the replication of ML processes.

\end{tabular}
\end{table}

\subsubsection{Data perspective:} 
\label{subsubsec:data_perspective}

Data is essential for ML-enabled systems. Poor data will result in inaccurate predictions, which is referred to in the ML context as “garbage in, garbage out”. Hence, ML requires high-quality input data. From the viewpoint of RE, it is clear that data constitutes a new type of requirements~\cite{challa2020faulty},~\cite{vogelsang2019requirements}. Based on the Data Quality model defined in the standard ISO/IEC 25012~\cite{iso25012} and our own experience, we elaborate on the data perspective. Table~\ref{table:data_properties} presents the data concerns and a brief description of each one.

\begin{table}[h]
\scriptsize
\centering
\caption{Data Perspective.}
\label{table:data_properties}
\begin{tabular}{|p{2.5cm}|p{9.5cm}|}
\hline
\multicolumn{1}{|c|}{\textbf{Concern}} & \multicolumn{1}{c|}{\textbf{Description (Addressing this concerns involves ... )}}\\ \hline

Accuracy         & Specifying the need to get correct data.\\ \hline

% how correctly the data represent the true value of the attributes.

Baseline       & Specifying the need for a baseline dataset approved by a domain expert that reflects the problem. It is employed to monitor other data acquired afterwards. \\ \hline

% Quality dataset approved by a domain expert that reflects the problem. It is employed to monitor other data acquired afterwards.

Bias             & Specifying the need to get data fair samples and representative distributions. Eventually this concern may also apply to the model perspective. \\ \hline

% disproportionate weight in favor or against something that results from an unfair sampling of a population.

Completeness     & Specifying the need to get data containing sufficient observations of all situations where the model will operate.\\ \hline

% the data shall contain observations of all the situations where the model will operate

Consistency      & Specifying the need to get consistent data in a specific context.\\ \hline

% the data shall be free from contradiction and be coherent with each other in a specific context.

Credibility      & Specifying the need to get true data that is believable and understandable by users.\\ \hline

% the data is regarded as true and believable by users.

Data operations      & Specifying what operations must be applied on the data (e.g., data cleaning and labeling). \\ \hline

% raw data must be converted into the representation the model needs. Operations such as data cleaning and data labeling can be applied.

Distribution mismatches   & Specifying expected data distributions and how data will be split into training and testing data.\\ \hline

% the data shall be split into some ratio of training and testing subsets.

\pbox{5cm}{Ethics and\\privacy}     & Specifying the need to get data to prevent adversely impacting society (e.g., listing potential adverse impacts to be avoided). Eventually this concern may also apply to the model perspective.\\ \hline

% it evaluates data practices, e.g., collecting,  analyzing and disseminating data, that have the potential to adversely impact society. 

Quantity         & Specifying the expected amount of data according to the type of the problem and the complexity of the algorithm. \\ \hline

% the amount of data depends on the complexity of the problem and on the complexity of the algorithm.

Real usage       & Specifying the need to get real data representing the real problem.\\ \hline

% the data has to be representative of the real world.

Source           & Specifying from where the data will be obtained. \\ \hline

% how and from where the data will be obtained.

Timeliness       & Specifying the time between when data is expected and when it is readily available for use.\\ \hline

% the time between when data is expected and when it is readily available for use.

\end{tabular}
\end{table}

%Our initial proposal considered a diagram with three requirements perspectives: model, data and system interaction. Each of these perspectives point out a set of concerns to be considered for specifying. The model perspective is concerned with properties that are in charge of data scientists such as explainability, inference time, learning time, ML metrics and reproducibility. Regarding the data perspective, we included data inherent properties (e.g., consistency, availability, bias, accuracy) should keep in mind to improve the chance of building suitable models. On the other hand, a system interaction perspective was considered. Within this, there is a user experience division in charge of properties such as visualization of model predictions, frequency and forcefulness of interactions and cost of mistakes. The other division is focused on infrastructure that needs to be put in place to support the operations a model requires. 

%We combine these elements in a diagram that provide a higher-level view of the ML-enabled system. With this, we represent the goals and concerns of each perspective, define and organize behaviors both functional and non-functional and specify the boundaries and requirements of an ML-enabled system. Due to their simplistic nature, diagrams can be a good communication tool for stakeholders. Figure~\ref{fig:initial_proposal} illustrates an overview of the initial proposal for specifying ML concerns.

\section{Focus Group}
\label{sec:focus_group}

%In this section, we present the focus group we conducted to evaluate our catalogue and gain feedback which contributed to improve it.

\subsection{Research Questions}
\label{subsec:focus_group_rq}

Based on our goal, we defined the following research question: Is our catalogue promising and could it support the requirements specification of ML-enabled systems? To answer this question, we evaluate this work from three angles. First, the perception of importance to know if the catalogue of concerns is addressing a relevant problem. Second, the perception of quality to establish if the catalogue of concerns is complete, consistent, correct and unambiguous, and third, the perception of feasibility to have an idea to what extent the catalogue of concerns can be applied in practice. For this purpose, we designed a focus group session for promoting in-depth discussion about the catalogue and its suitability. Focus group is a qualitative research method based on gathering data through the conduction of group interviews and it has been conducted in SE for revealing arguments and feedback from practitioners~\cite{kontio2004using}.

\subsection{The Participants}
\label{subsec:focus_group_participants}

We invited eight practitioners who have been actively working with the development of ML-enabled systems in industry. Before conducting the focus group session, we applied a characterization form. We asked them about the role they perform within their company, and their experience in years and number of ML projects they participated in. Table~\ref{table:overview_participants} shows an overview of the participants.  

\begin{table}[h]
\scriptsize
\centering
\caption{Overview of the participants}
\label{table:overview_participants}
\begin{tabular}{|l|l|c|c|}
\hline
\multicolumn{1}{|c|}{\textbf{Id}} & \multicolumn{1}{c|}{\textbf{Role}} & 
\textbf{\# years} & \textbf{\# ML projects} \\ \hline
P1                                & Data scientist                     & 13                & 12                   \\ \hline
P2                                & Data scientist                     & 9                 & 7                    \\ \hline
P3                                & Data scientist                     & 1                 & 3                    \\ \hline
P4                                & Developer                          & 1                 & 1                    \\ \hline
P5                                & Developer                          & 3                 & 2                    \\ \hline
P6                                & Project lead                       & 1                 & 2                    \\ \hline
P7                                & Project lead                       & 2                 & 5                    \\ \hline
P8                                & Project lead                       & 2                 & 2                    \\ \hline
\end{tabular}
\end{table}

\subsection{Execution}
\label{subsec:focus_group_instrumentation_execution}

Before starting the focus group, we introduced to the participants the main challenges when engineering ML-enabled systems and presented how RE may address some of them. The focus group was conducted in a Zoom meeting recorded for the study. The study was planned to be executed in two phases. In the first phase that took 20 minutes, we explained our catalogue by decomposing each requirement perspective in its related concerns. The second phase, that took 45 minutes, was a question and answer session about the importance, quality and feasibility of our catalogue of concerns. The focus group was moderated by the first and second author and was recorded and transcribed.

\subsection{Results}
\label{subsec:focus_group_results}

%After the focus group session, the content of the recordings and the notes was analyzed by two researchers. In the following, we present the results.

\subsubsection{Perception of Importance:}
\label{subsubsec:importance}

We asked the participants how they define, document and organize requirements for ML-enabled systems and if they think it is important. P7 stressed the lack of formal methods to support their definition, modeling and documentation: “I have constant difficulties to find tools and methods to help my team and customers understand ML requirements”. On the other hand, P2 stated: “In my opinion, the requirements process for ML is ad-hoc, which makes it highly dependent on people’s knowledge” and P3 manifested: “I noticed that requirements have constant rework in ML projects”. Considering the overall discussion, we understood that creating new methods in this direction is absolutely important to address the problems practitioners are facing.

\subsubsection{Perception of Quality:}
\label{subsubsec:quality}

The participants evaluated our catalogue by (i) analyzing the concerns and perspectives in terms of completeness, consistency, correctness and ambiguity, and by (ii) measuring the capacity of our catalogue to support the specification of ML-enabled systems. Overall, there was a clear consensus that practitioners are unaware of the big picture. They did not know about many of the concerns, while judging them as relevant and helpful to support more precise specifications. P1 mentioned that "I wish I had such concerns specified upfront in my ML projects, decisions regarding these concerns should not be taken without appropriately involving stakeholders or when coding". When analyzing the perspectives, P2 emphasized the importance of the ML objective perspective: “I understand the need to consider data, model, user and infrastructure, but in my opinion, the functional behaviour of ML models, which is reflected in the objectives, is crucial”. Regarding the concerns, P5 stated that from the technical view, the concerns and their grouping make sense: “When seeing the concerns I was able to relate them to problems and tasks I faced in the past”. P1 manifested the importance to evaluate in depth the completeness of the concerns. For instance: “In the data perspective, a common concern is the definition of a baseline that helps in the acquisition of new data. I would definitely consider it”. We improved the catalog based on the session feedback. 

\subsubsection{Perception of Feasibility:}
\label{subsubsec:feasibility}

The participants found the catalogue of concerns useful to support the requirements specification of ML-enabled systems. P6 stressed the concerns organization into perspectives: “I think the catalogue can help us to analyze ML requirements since it covers several perspectives for different situations”. In addition, P8 stated: “We need to use this type of proposals in practice due to the overview that it provides and the concerns that may apply in our context”. Hence, we understood that it is feasible to further evaluate our catalogue practical contexts.

\section{Discussion}
\label{sec:discussion}

%Developing ML-enabled systems involves a set of skills reflected in three areas: data engineering, data science and SE. Our perception is that many companies have data engineers writing REST APIs, data scientists building pipelines and software engineers building models. This can lead to extra efforts and low software quality. In addition, we see that ML practitioners often scribble a few Jupyter Notebooks to build and evaluate models where code quality is bad. They run experiments and the artifacts generated are saved to folders named in mysterious ways, randomly spread across the filesystem. On the other hand, documentation is missing and the implementation is difficult to understand. In the end, ML practitioners identify the best performing model and then they deploy it into production. This leads to redo substantial work. That's where technical debt comes in. 

%Many companies have data engineers writing REST APIs, data scientists building pipelines and software engineers building models. In ML projects, there is a perception that time and money is wasted even to a greater extent than in non-ML projects. Our catalogue seeks to avoid pushing ML practitioners into inefficient generalist tasks that do not address relevant issues. 

We are aware that not every ML-enabled system needs to address all the concerns we proposed and not every ML-enabled system needs to implement them to the same degree. Our intention is to provide an overview of concerns so that requirements engineers can analyze the needs of their ML-enabled systems with stakeholders. 

It is noteworthy that this overview focuses on concerns related to ML-enabled part of ML-enabled systems and the integration of this part with the remainder of the system. However, when considering the overall system, general quality characteristics of software products such the ones mentioned in the ISO/IEC 25010 standard~\cite{iso25010}, should also be analyzed. 

The main contribution of our work is to provide a set of concerns grouped into perspectives that can be used by requirements engineers to support the specification of ML-enabled systems. Nevertheless, we believe our work may eventually be useful in various situations. First, to validate an already specified system. In this case, our concerns would be a reference since they come from a literature review and different industrial experiences on building ML-enabled systems. Second, our work helps to understand ML modularity since it provides several components and details at different levels about functional and non-functional aspects. Third, it is applicable to the most common ML approaches. Our work is focused on supervised and unsupervised ML problems. Both of them learn from data. Therefore, ML-enabled system would benefit from, at least, analyzing the perspectives and concerns we proposed.
\section{Concluding Remarks}
\label{sec:conclusions}

The development of ML-enabled systems involves understanding business context and problems, translating them into ML tasks, designing and experimenting with algorithms, evaluating models, designing pipelines, among other tasks. This needs to be considered from early stages of ML software development. Based on the literature and on practical experiences, we proposed a catalogue of 45 concerns to support the specification of ML-enabled systems that covers five perspectives: objectives, user experience, infrastructure, model, and data. This is the first effort aiming at providing the big picture of concerns for specifying ML-enabled systems. With this catalogue, we seek to empower requirements engineers with an ML overview of concerns that should be analyzed together with business owners, data scientists, software engineers and designers. 

We evaluated the catalogue by conducting a focus group session with eight ML practitioners involved in developing ML-enabled systems. The purpose was to gain insights about the relevance of the problem we are addressing, the benefits of using it and its feasibility. The results indicated that practitioners consider the identified concerns relevant and the catalogue useful. They stated that grouping perspectives and organizing concerns can help them to identify constrains upfront with other practitioners. Therefore, we believe that the conceptual perspectives and concerns we herein proposed can be helpful to support specifying ML-enabled systems. Future work includes conducting additional evaluations in industry settings, including case studies.

%\input{Sections/Example.tex}

\bibliographystyle{splncs04}
\bibliography{bibTex/references}
\end{document}